\documentclass{article}

\usepackage[utf8]{inputenc} % allow utf-8 input
\usepackage[T1]{fontenc}    % use 8-bit T1 fonts
\usepackage{hyperref}       % hyperlinks
\usepackage{amsmath}        % math environments and \text in math mode
\usepackage{cleveref}       % smart cross-referencing (provides \Cref)
\usepackage{url}            % simple URL typesetting
\usepackage{booktabs}       % professional-quality tables
\usepackage{amsfonts}       % blackboard math symbols
\usepackage{nicefrac}       % compact symbols for 1/2, etc.
\usepackage{microtype}      % microtypography
\usepackage[dvipsnames]{xcolor}         % colors
\usepackage{graphicx}      % includegraphics for figures
\usepackage{subcaption}    % subfigure support
\usepackage{wrapfig}

\usepackage[preprint]{neurips_2026}

\usepackage{amsmath}
\usepackage{bm}
\usepackage{booktabs}
\usepackage{multirow}
\usepackage{tikz}
\usetikzlibrary{positioning,arrows.meta,fit,backgrounds}
\usepackage{pgfplots}
\pgfplotsset{compat=1.17}
\usepgfplotslibrary{groupplots}
\usepackage{xcolor}

\definecolor{palBlue}{HTML}{7AA4CA}
\definecolor{palOrange}{HTML}{F3B06F}
\definecolor{palGreen}{HTML}{A3B478}
\definecolor{palPurple}{HTML}{A894C1}
\definecolor{palRed}{HTML}{D6897C}
\definecolor{palTeal}{HTML}{74B2AF}
\definecolor{palPink}{HTML}{E6B8D2}
\colorlet{cComposite}{palTeal}
\colorlet{cBio}{palBlue}
\colorlet{cBatch}{palOrange}

\newcommand{\mcD}{\mathcal{D}}

\newcommand{\mcX}{\mathcal{X}}
\newcommand{\mcZ}{\mathcal{Z}}

\newcommand{\MMD}{\mathrm{MMD}}
\newcommand{\bx}{\bm{x}}
\newcommand{\by}{\bm{y}}
\newcommand{\bA}{\bm{A}}
\newcommand{\bb}{\bm{b}}
\newcommand{\ba}{\bm{a}}

\newcommand{\diag}{\mathrm{diag}}

\begin{document}

\title{Confounder-Aware Feature Correction for\\ Single-Cell Batch Integration}

\author{%
  Calvin McCarter\\
  \texttt{mccarter.calvin@gmail.com}%
}

\maketitle

\begin{abstract}
Batch integration is a central preprocessing step in single-cell genomics, where datasets collected across experiments, donors, and protocols must be combined despite pervasive technical batch effects. The leading integration methods produce a shared low-dimensional embedding, which discards the corrected gene-expression values that downstream differential-expression, biomarker, and other analyses depend on. The feature-space (expression-correcting) methods that do preserve genes typically align the \emph{marginal} expression distributions of batches and thereby risk erasing genuine biological variation whenever cell-type composition differs across batches---i.e.\ whenever batch effect and biological signal are \emph{confounded}. We recast single-cell batch integration as confounded domain adaptation and apply ConDo, a method that matches \emph{conditional} expression distributions given the cell-type annotation rather than marginal distributions. To extend ConDo's pairwise source-to-target adapter to the many-batch setting, we introduce an agglomerative compatibility-graph integrator: batches are nodes connected when they share a cell type, and we greedily merge each best-scoring neighbor into a growing reference by fitting one ConDo adapter. On the Open Problems Batch Integration Benchmark, ConDo is the strongest feature-space integrator, ranking first among feature methods on five of six datasets. Furthermore, it is competitive with or better than deep embedding methods on the overall score, ranking first across all methods on four of six datasets while returning corrected expression rather than an opaque embedding.
\end{abstract}

\section{Introduction}
\label{sec:intro}

Single-cell RNA sequencing (scRNA-seq) atlases are now routinely assembled from many experiments spanning donors, tissues, laboratories, and chemistries. Combining these data requires correcting ``batch effects'': technical differences that shift the measured expression of otherwise identical cells \citep{leek2010tackling}. The difficulty is that batch effects are rarely the only thing that differs between batches. Different experiments also sample different mixtures of cell types and states, so the observed shift between two batches mixes a technical component (to be removed) with a biological component (to be preserved). When these two are statistically entangled we say batch and biology are \emph{confounded}, and naive alignment that forces two batches to look identical will discard real biological structure along with the technical artifact.

Integration methods differ in what they return, and this determines what they can support downstream. The dominant family of methods sidesteps the confounding problem by returning a shared low-dimensional embedding in which batches overlap, most notably Harmony \citep{korsunsky2019fast}, scVI \citep{lopez2018deep}, scANVI \citep{xu2021probabilistic}, scPoli \citep{de2023population}, and recent foundation models \citep{cui2024scgpt,theodoris2023transfer,rosen2023universal}.
%--- or, in the case of BBKNN \citep{polanski2020bbknn}, a batch-balanced neighbor graph. 
These outputs serve visualization, clustering, and label transfer, but their coordinates are not gene expression: they cannot support marker discovery, co-expression analysis, gene-set scoring, or the many per-gene pipelines and pretrained models that expect a cell-by-gene matrix. Latent-variable models can decode to gene space \citep{boyeau2019deep}, but the result is a model-generated quantity inheriting the decoder's parametric assumptions rather than a corrected measurement. This consideration, together with the confounding problem above, gives four criteria for a batch correction method. It should (i) return a corrected expression matrix rather than an embedding; (ii) target an estimand stated in advance, so that its behavior under confounding follows from the target rather than from hyperparameter tuning; (iii) apply uniformly to every cell, including those whose type is unknown; and (iv) identify that target from the data at hand, without requiring a control arm that atlas-scale collections lack. No existing method satisfies all four.

Feature-space methods do return a matrix, but most demand the wrong alignment. The classical approach aligns each gene's \emph{marginal} distribution across batches, shifting and scaling toward a per-batch consensus \citep{johnson2007adjusting,ritchie2015limma,shaham2017removal}. Confounding makes this untenable: if batch~A is enriched for a cell type that batch~B lacks, the two marginals ought to differ, and forcing them to coincide removes that compositional difference. Adversarial and penalty-based autoencoders acknowledge the danger and add a countervailing term, trading a batch-mixing penalty against reconstruction loss \citep{wang2021imap,shree2023scdreamer}. This improves matters in practice but not in principle. Reconstruction is not biology preservation; it resists all movement of the data equally, and cannot distinguish a genuine biological difference from a technical artifact the model happens to fit poorly. Two objectives, each incorrect in isolation, are balanced in the hope that their errors offset.

Another set of feature-space methods avoids marginal alignment by working locally. Mutual nearest neighbor (MNN) correction pairs cells across batches and derives cell-specific corrections from the paired residuals \citep{haghverdi2018batch}; Scanorama extends this to partially overlapping collections \citep{hie2019efficient}, Seurat's anchors follow a related strategy \citep{stuart2019comprehensive}, and SCALEX learns a batch-free encoder \citep{xiong2022online}. These methods were designed with confounding in mind and explicitly relax the equal-composition assumption: a population present in one batch and absent from another is left uncorrected rather than forcibly aligned. Their limitation is different in kind. Each is defined by a procedure rather than an estimand: the correction is whatever the matching and averaging steps produce, with no population quantity of which it is an estimate. This is criterion (ii) failing in a different way from the marginal methods: not a wrong target but none at all. The conditions under which a mutual pair reflects the same biological state, rather than the nearest available substitute for an absent one, are neither stated nor checkable from the data, so behavior under confounding can only be assessed empirically, case by case. Results also depend on various hyperparameters, none of which carries an interpretation in terms of biological or technical variation. 
%Empirical benchmarks \citep{luecken2022benchmarking} are therefore the primary evidence available for these methods, and benchmark rankings are known to shift with dataset, preprocessing, and metric weighting.

Another family brings biology into the correction itself. ComBat admits a design matrix of covariates to be preserved, and scANVI \citep{xu2021probabilistic} and scGen \citep{lotfollahi2019scgen} use cell-type labels to constrain what is aligned. However, while conditioning is a fruitful avenue for addressing confounding, the manner of conditioning matters. When a label enters as a covariate of the correction, the fitted transformation becomes cell-type-dependent, giving each stratum its own map from batch~A to batch~B. We regard this as the wrong object to estimate. For example, when differences in sequencing depth act on the assay, the natural model of its effect is a single transformation applied to every cell irrespective of type; if a batch effect genuinely acts differently across cell types, the batches are not comparable in a way that any correction repairs, and the correct response is to decline to combine batches rather than to fit a more flexible map. Per-stratum maps also multiply free parameters and overfit the rare types where cells are scarce, and they violate criterion (iii) outright. 

CellANOVA \citep{zhang2025recovery} satisfies criteria (i)--(iii) and is the closest existing method to what the criteria describe. It defines batch effect as the variation remaining between control samples after conditioning on latent cell state, estimates a batch subspace from a designated pool of control samples, and projects that subspace out of every cell, returning a corrected expression matrix. Its identification, however, comes from experimental design: the analyst must nominate control samples known a priori to be free of the differences the study is about. Such designs exist in controlled case--control and longitudinal studies, but not in the atlas setting described at the outset, where data are assembled post hoc from independently conducted experiments with no common control arm; thus criterion (iv) is not met. 

We instead start from the observation that nothing in the conditional formulation requires the correction itself to depend on the label. A method could instead use labels only to \emph{identify} a single shared transformation, by requiring that it match the conditional distributions of expression given cell type; once identified on labeled cells, that transformation would apply to every cell, regardless of labeling.
This is precisely the setting addressed by Confounded Domain Adaptation (ConDo) \citep{mccarter2024towards}. ConDo learns a feature-space transformation that matches the conditional distribution of features given a confounder, rather than the marginal distribution of features. In the typical single-cell scenario, the confounder is the cell-type annotation: ConDo aligns the expression of cells of the same cell type across batches, while leaving compositional differences between cell types intact. 
ConDo can also accommodate multiple cell annotations and even quantitative annotations by fitting conditional generative models of expression given biological confounders, but in this work we focus on using the standard cell-type annotations.

ConDo as originally formulated is a pairwise adapter: it maps one source domain onto one target domain. Single-cell integration usually involves tens of batches with partially overlapping cell-type compositions, so a pairwise adapter does not directly apply. Our main methodological contribution is an agglomerative compatibility-graph integrator that lifts ConDo to the many-batch setting. We build a graph whose nodes are batches and whose edges connect batches that share at least one cell type; starting from the batch whose cells are best resolved by type, we repeatedly select the highest-scoring batch adjacent to the current integrated pool and fit a single ConDo adapter mapping it onto that pool, growing the reference at every step. This guarantees that every adapter is fit between batches with overlapping cell types (so the conditional matching is well posed) and that later batches are aligned to a large, diverse reference rather than a single small batch.

On the Open Problems Batch Integration Benchmark \citep{openproblems2025}, we show that ConDo is the strongest feature-space integrator---first among feature methods on five of six atlases---and is competitive with or better than state-of-the-art deep embedding methods on the overall SCIB score, while uniquely returning corrected expression (Section~\ref{sec:experiments}). We further show that ConDo is robust to missing cell-type labels: excluding unlabeled cells from the fit preserves the corrected output across all six atlases even when most labels are withheld (Section~\ref{sec:results-dropout}).
Because cell-type annotation is a required field of the benchmark's method input (Section~\ref{sec:setup}), ConDo is, like scANVI, a label-aware integrator; we make this distinction explicit throughout and compare most directly against scANVI.

Code is provided at \url{https://github.com/calvinmccarter/condo-adapter}.

\iffalse
Our contributions are:
\begin{itemize}
\item We recast single-cell batch integration as confounded domain adaptation with the cell-type annotation as confounder, yielding a feature-space integrator that corrects technical shift without erasing confounded biological variation (Section~\ref{sec:method-condo}).
\item We introduce the agglomerative compatibility-graph integrator, a general procedure for applying any pairwise conditional-distribution-matching adapter to the many-batch setting (Section~\ref{sec:method-agg}).
\item
\end{itemize}
%We position ConDo against prior integration and domain-adaptation methods in Appendix~\ref{app:related}.
\fi

\section{Background}
\label{sec:background}

\subsection{Confounded shift and ConDo}
\label{sec:bg-condo}

We summarize the parts of ConDo \citep{mccarter2024towards} needed here. Let $\mcX$ be feature (expression) space and $\mcZ$ a confounder space; a domain is a joint distribution over $\mcX \times \mcZ$, and we have a source $\mcD_S$ and target $\mcD_T$ with marginals $\mcD^X,\mcD^Z$. ConDo addresses \emph{confounded shift}: both marginals may differ across domains ($\mcD_S^X \neq \mcD_T^X$, $\mcD_S^Z \neq \mcD_T^Z$), but there exists a feature-space map $g:\mcX_T \to \mcX_S$ matching the conditionals,
\begin{equation}
\mcD_S(X \mid Z=z) = \mcD_T\big(g(X) \mid Z=z\big) \qquad \forall z .
\label{eq:confounded-shift}
\end{equation}
Crucially $g$ takes features only---not $z$---so it applies to cells whose confounder is unknown and yields general-purpose corrected features; this distinguishes ConDo from ComBat, whose per-confounder term requires $z$ at correction time. ConDo fits $g$ by minimizing the expected conditional divergence under a prior $\hat{\mcD}^Z_\times$ over the confounder,
\begin{equation}
\min_{g} \; \mathbb{E}_{z \sim \hat{\mcD}^Z_\times} \, d\Big(\mcD_T(\bx \mid Z=z),\; \mcD_S(g(\bx)\mid Z=z)\Big),
\label{eq:framework}
\end{equation}
where $\hat{\mcD}^Z_\times$ estimates the product of the source and target confounder distributions, weighting confounder values well supported in both domains (so both conditionals are reliably estimable). For a categorical confounder this is just a normalized product of per-value source and target frequencies.

Conditional-distribution matching across domains has been studied in the domain adaptation literature as generalized label shift and target/conditional shift \citep{tachet2020domain,zhang2013domain}, but those methods reweight samples or optimize a single downstream label rather than producing reusable corrected features as does ConDo.

\subsection{Restricting the transformation}
\label{sec:bg-transform}

We restrict $g$ to be linear, $g(\bx) = \bA\bx + \bb$, in two forms that encode different assumptions about the batch effect. \emph{Location-scale} takes $\bA=\diag(\ba)$, correcting each gene by its own shift and scale, $[g(\bx)]_i = a_ix_i+b_i$ (as ComBat does), and assumes the technical effect acts per gene. \emph{Affine} takes $\bA$ dense, letting the correction mix genes to undo effects that rotate or shear the expression space (e.g.\ a protocol change re-weighting co-expressed gene programs), at the cost of $O(d^2)$ parameters. These are not a quality/speed tradeoff but two distinct modeling choices. 
The appropriate choice depends on the technical effect in a given dataset (Section~\ref{sec:results-variants}); 
The location-scale option may also be preferred in certain scenarios, due to its interpretability, its preservation of correlation structure, or its guarantee that the post-correction expression value for a gene is solely determined by the pre-correction expression value for that gene. 

\subsection{The conditional MMD objective}
\label{sec:bg-mmd}

We take $d$ in Eq.~\eqref{eq:framework} to be the squared maximum mean discrepancy \citep{gretton2012kernel} under a characteristic kernel $k_\mcX$ (RBF). At a confounder value $z$, with $g(\bx)=\bA\bx+\bb$,
\begin{align}
\MMD^2\big(z\big) =\;&
\mathbb{E}_{\bx,\bx' \sim \mcD_T(\cdot\mid z)} k_\mcX(\bx,\bx')
- 2\,\mathbb{E}_{\substack{\bx \sim \mcD_T(\cdot\mid z)\\ \by \sim \mcD_S(\cdot \mid z)}} k_\mcX\big(\bx,\, \bA\by+\bb\big) \nonumber\\
&+ \mathbb{E}_{\by,\by' \sim \mcD_S(\cdot \mid z)} k_\mcX\big(\bA\by+\bb,\, \bA\by'+\bb\big).
\label{eq:condo-mmd}
\end{align}
We minimize $\mathbb{E}_{z\sim\hat{\mcD}^Z_\times}\MMD^2(z)$ by stochastic gradient descent with AdamW \citep{loshchilov2017fixing} optimizer. Because the cell-type confounder is categorical, sampling from $\mcD_\cdot(\bx\mid Z=z)$ is trivial: we draw cells of type $z$ from the relevant batch.

\section{Method}
\label{sec:method}

\subsection{ConDo for a pair of batches}
\label{sec:method-condo}

Given two batches with normalized expression $\bx$ and cell-type labels, we treat one as source $\mcD_S$ and one as target $\mcD_T$, set the confounder $Z$ to the cell-type label, and fit $g(\bx)=\bA\bx+\bb$ by minimizing the conditional-MMD objective of Eq.~\eqref{eq:condo-mmd} under the product-prior weighting over shared cell types. The fitted $g$ maps source cells onto the target's expression distribution \emph{cell-type by cell-type}: cells of a type present in both batches are aligned, while the relative abundances of cell types are untouched, so a cell type abundant in one batch and rare in the other is not artificially equalized. Algorithm hyperparameters (such as batch size and AdamW parameters) follow the ConDo defaults, with the same hyperparameters used across all experiments.

\subsection{Agglomerative compatibility-graph integration}
\label{sec:method-agg}

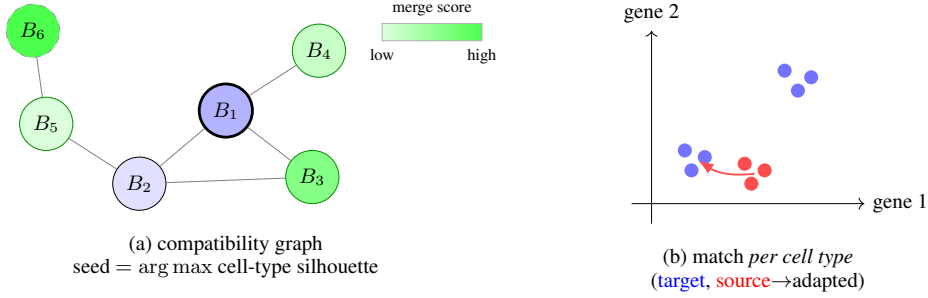
\begin{figure}[t]
\scalebox{0.88}{
\begin{tikzpicture}[
  font=\small,
  batch/.style={circle,draw,minimum size=8mm,inner sep=1pt},
  pool/.style={batch,fill=blue!12},
  seed/.style={batch,fill=blue!30,very thick},
  cand/.style={batch,fill=green!#1,draw=green!55!black},
  cand/.default=20,
  blocked/.style={draw=gray!80,dashed},
  edge/.style={-,gray},
  merge/.style={-{Latex},thick}
]
% left: compatibility graph (indices = merge order)
\node[seed]  (b1) at (0,0)        {$B_1$};   % seed
\node[pool]  (b2) at (-1.3,-1.1)  {$B_2$};   % already merged
\node[cand=48] (b3) at (1.3,-1.0) {$B_3$};   % best eligible -> merged next
\node[cand=22] (b4) at (1.4,0.9)  {$B_4$};
\node[cand=14] (b5) at (-2.7,-0.2){$B_5$};
\node[cand=70,blocked] (b6) at (-2.9,1.2) {$B_6$}; % top score, but not adjacent to pool
\draw[edge] (b1)--(b2); \draw[edge] (b1)--(b3); \draw[edge] (b1)--(b4);
\draw[edge] (b2)--(b3); \draw[edge] (b2)--(b5);
\draw[edge] (b5)--(b6);
% score colorbar
\begin{scope}[shift={(2.35,1.05)}]
  \shade[left color=green!12,right color=green!70,draw=gray!60]
        (0,0) rectangle (1.5,0.26);
  \node[font=\scriptsize,anchor=south] at (0.75,0.26) {merge score};
  \node[font=\scriptsize,anchor=north] at (0,0)   {low};
  \node[font=\scriptsize,anchor=north] at (1.5,0) {high};
\end{scope}
\node[align=center] at (0,-2.2) {(a) compatibility graph\\seed $=\arg\max$ cell-type silhouette};

% right: schematic of conditional vs marginal matching
\begin{scope}[xshift=6.4cm,yshift=-1.4cm]
\draw[->] (-0.3,0) -- (3.2,0) node[right]{gene 1};
\draw[->] (0,-0.3) -- (0,2.6) node[above]{gene 2};
% target clusters (two cell types)
\fill[blue!55] (0.6,0.5) circle (3pt); \fill[blue!55] (0.8,0.7) circle (3pt); \fill[blue!55] (0.5,0.8) circle(3pt);
\fill[blue!55] (2.2,1.7) circle (3pt); \fill[blue!55] (2.4,1.9) circle (3pt); \fill[blue!55] (2.0,2.0) circle(3pt);
% source clusters shifted (one cell type only)
\fill[red!70] (1.5,0.3) circle (3pt); \fill[red!70] (1.7,0.5) circle (3pt); \fill[red!70] (1.4,0.6) circle(3pt);
\draw[merge,red!70] (1.55,0.45) to[bend left=20] (0.7,0.65);
\node[align=center] at (1.6,-1.0) {(b) match \emph{per cell type}\\(\textcolor{blue}{target},\ \textcolor{red}{source}$\to$adapted)};
\end{scope}
\end{tikzpicture}
}
\caption{(a) The agglomerative integrator seeds at the batch with the highest within-batch cell-type silhouette and greedily merges the best-scoring graph neighbor of the current pool, fitting one ConDo adapter per merge. Indices give the merge order. \textcolor{blue!70!black}{Blue} nodes are batches already in the reference pool ($B_1$, thickly outlined, is the seed); \textcolor{green!50!black}{green} nodes are unmerged batches, shaded by merge score (darker $=$ higher). Only batches adjacent to the pool are eligible, so $B_3$ is merged next even though the dashed $B_6$ scores highest: $B_6$ shares no edge with the pool and is merged last, becoming eligible only once $B_5$ is absorbed. (b) Each adapter matches source to target \emph{conditional on cell type}, so a cell type present in only one batch (here the lower-left cluster) is aligned without forcing the batches' marginal distributions to coincide.}

\label{fig:schematic}
\end{figure}

A real dataset has $B$ batches, not necessarily two; the final result depends on the order in which pairwise adaptation is applied, so the ordering must be chosen carefully. Furthermore, not every pair of batches shares cell types, and fitting an adapter between two batches with \emph{no} shared cell type is ill posed, because the product prior of Eq.~\eqref{eq:framework} has empty support. We address both issues with an agglomerative integrator (Figure~\ref{fig:schematic}).

\paragraph{Compatibility graph.} We build an undirected graph $G$ with one node per batch and an edge between two batches iff their cell-type sets intersect. Edges mark exactly the pairs for which a ConDo adapter is well posed. We assume $G$ has a single connected component, which holds for the atlas datasets we consider since their batches almost always share common cell types.%; we return to the disconnected case in Section~\ref{sec:discussion}.

\paragraph{Seed and scoring.} We score each batch by the silhouette coefficient \citep{rousseeuw1987silhouettes} of its cell-type labels on the precomputed PCA---a label-aware measure of how cleanly cell types separate within that batch, computable before any integration. The seed (initial reference) is the highest-scoring batch, so integration is anchored to the batch whose biology is internally best resolved.

\paragraph{Greedy growth.} We maintain a target pool, initialized to the seed batch. At each step, we consider all batches adjacent in $G$ to any batch already in the pool, pick the highest-scoring such batch, fit one ConDo adapter mapping that batch (source) onto the current pool (target) conditioned on cell type, replace the batch's cells with their adapted values, and add it to the pool. We repeat until no adjacent batch remains. Because the pool grows, later batches are aligned to an increasingly cell-type-diverse reference, rather than every batch being adapted to a single (possibly small) batch. Appendix~\ref{app:ordering} ablates the scoring rule, showing that ConDo's performance is robust to sensible merge orders.

\section{Experiments}
\label{sec:experiments}

\subsection{Setup}
\label{sec:setup}

\begin{table}[h]
\centering
\caption{Benchmark datasets in the Open Problems Batch Integration Benchmark. Cell and batch counts are approximate.}
\label{tab:datasets}
\begin{tabular}{lrrl}
\toprule
Dataset & Cells & Batches & Organism / tissue \\
\midrule
\textit{DKD}                & 39k  & 11 & human kidney (diabetic kidney disease) \\
\textit{GTEx}             & 209k & 16 & human, GTEx multi-tissue \\
\textit{ImmuneCell}    & 330k & 12 & human immune cells \\
\textit{MousePancreas} & 302k & 56 & mouse pancreas \\
\textit{Hypomap}            & 385k & 24 & mouse hypothalamus \\
\textit{TabulaSapiens}      & 483k & 29 & human, multi-organ \\
\bottomrule
\end{tabular}
\end{table}

We use six human and mouse atlases from the Open Problems Batch Integration Benchmark \citep{openproblems2025} (Table~\ref{tab:datasets}), ranging from 39k to 483k cells and 11 to 56 batches. Each method receives the normalized expression matrix, per-cell \texttt{batch} and \texttt{cell\_type} annotations, and a precomputed 50-dimensional PCA; \texttt{cell\_type} is a required input field of the benchmark. ConDo and scANVI use the cell-type annotation; the other baselines (ComBat, Harmony, scVI, Scanorama, SCALEX, fastMNN, LIGER, scGPT, Geneformer, and others) use only \texttt{batch}. We run two ConDo variants, MMD-affine and MMD-location-scale, on the normalized expression (feature output). We evaluate on the Open Problems Batch Integration Benchmark (OPBIB), the maintained successor to the Single-Cell Integration Benchmark (SCIB) benchmark \citep{luecken2022benchmarking}, following its evaluation exactly: feature outputs are PCA-reduced to a 50-D embedding on which the SCIB metrics are computed, matched to the official per-metric scripts. The metrics split into \emph{bio-conservation} (cell-type ASW, NMI, ARI, isolated-label ASW/F1, cLISI, HVG overlap, cell-cycle conservation) and \emph{batch-correction} (batch ASW, graph connectivity, PCR, iLISI, kBET). Each metric is min-max scaled across the full method pool (including the no-integration, shuffle, and oracle controls that anchor the scale) before averaging, and the overall score is $0.6\,\text{bio}+0.4\,\text{batch}$. Our numbers use a development snapshot of the benchmark's public baselines (the six-atlas \texttt{cellxgene\_census} generation; the later v2.0.0 release additionally scores cell-cycle conservation on all six atlases). Three SCIB metrics are not uniformly available in this snapshot: cell-cycle conservation (computed on only three atlases), isolated-label F1 (undefined on \textit{DKD}), and kBET (missing for many method--dataset pairs). Following the SCIB methodology we drop such non-uniform metrics.% and average the rest.
%Appendix~\ref{app:webscore} reports the alternative equal-weight aggregation the current \texttt{openproblems.bio} leaderboard uses instead.

We report two leaderboards. The \emph{feature} leaderboard compares only expression-correcting methods, on the largest metric set they share (10 metrics including HVG overlap, cell-type ASW, and PCR, which embedding methods structurally lack). The \emph{all-methods} leaderboard compares every method---feature and embedding---on the 7 metrics common to both. Because location-scale and affine encode different assumptions, we also report a \emph{best-of-two} configuration that selects, per dataset, whichever transform scores higher; we discuss how this selection can be made in practice in Section~\ref{sec:results-variants}.

\subsection{ConDo is the strongest feature-space integrator}
\label{sec:results-feature}

Table~\ref{tab:feature} compares ConDo against the other expression-correcting methods. The best-of-two ConDo configuration ranks first among feature methods on five of the six atlases, trailing only ComBat on \textit{GTEx}. Averaged across datasets, ConDo attains mean rank $1.17$ versus ComBat's $1.83$ within the two-method full-coverage pool; its mean shortfall from the best feature method per dataset is only $1.0$ overall-score point, versus ComBat's $5.3$ (Scanorama, SCALEX, and MNN-based methods are far behind). The gap is concentrated in bio-conservation, consistent with the hypothesis that conditioning on cell type preserves biological structure that marginal alignment removes.

\begin{table}[h]
\centering
\small
\setlength{\tabcolsep}{3.5pt}
\caption{Feature-space methods: overall SCIB score per dataset (higher is better), feature leaderboard (10-metric feature set, which adds HVG overlap, cell-type ASW, and PCR to the 7 metrics of Table~\ref{tab:all}; scores are therefore not directly comparable across the two tables). \textbf{Bold} marks the best feature method per dataset. ConDo (best-of-two) is first on 5/6 datasets. Dashes mark methods that failed to produce a scoreable output on a dataset.}
\label{tab:feature}
\begin{tabular}{lcccccc}
\toprule
Method & DKD & GTEx & Hypomap & Immune & Mouse & TabSap \\
\midrule
\textbf{ConDo (best-of-two)} & \textbf{0.815} & 0.646 & \textbf{0.715} & \textbf{0.807} & \textbf{0.781} & \textbf{0.702} \\
\quad ConDo affine          & 0.815 & 0.646 & 0.715 & 0.794 & 0.670 & 0.702 \\
\quad ConDo location-scale  & 0.805 & 0.555 & 0.679 & 0.807 & 0.781 & 0.516 \\
ComBat                          & 0.764 & \textbf{0.705} & 0.613 & 0.734 & 0.720 & 0.673 \\
SCALEX                          & 0.623 & 0.532 & 0.518 & 0.554 & --    & 0.549 \\
MNN                             & 0.681 & --    & --    & --    & --    & --    \\
Scanorama                       & 0.323 & 0.199 & 0.204 & 0.131 & 0.252 & --    \\
mnnpy                           & 0.320 & --    & --    & --    & --    & --    \\
\bottomrule
\end{tabular}
\end{table}
\begin{table}[h]
\centering
\small
\setlength{\tabcolsep}{4.5pt}
\caption{All methods, overall SCIB score per dataset (all-methods leaderboard, 7 metrics common to feature and embedding methods; not directly comparable to the 10-metric Table~\ref{tab:feature}). \textbf{Bold} marks the best method per dataset; $^\ast$ marks label-aware methods (ConDo, scANVI). ConDo returns corrected expression; the embedding methods do not.}
%\caption{Single-Cell Integration Benchmarking (SCIB) score per dataset (all-methods leaderboard, 7 metrics common to feature and embedding methods; \textbf{Bold} marks the best method per dataset; $^\ast$ marks cell-type-aware methods (ConDo, scANVI). ConDo returns corrected expression; the embedding methods do not.}
\label{tab:all}
\setlength{\tabcolsep}{3.5pt}
\begin{tabular}{llcccccc}
\toprule
Method & type & DKD & GTEx & Hypomap & Immune & Mouse & TabSap \\
\midrule
\textbf{ConDo (best-of-two)} & feat$^\ast$ & \textbf{0.830} & 0.695 & \textbf{0.612} & 0.760 & \textbf{0.770} & \textbf{0.667} \\
\quad ConDo affine          &             & 0.830 & 0.695 & 0.612 & 0.760 & 0.701 & 0.667 \\
\quad ConDo location-scale  &             & 0.787 & 0.602 & 0.539 & 0.750 & 0.770 & 0.528 \\
scANVI            & emb$^\ast$ & 0.801 & \textbf{0.742} & 0.499 & \textbf{0.766} & 0.763 & 0.633 \\
scVI              & emb. & 0.781 & 0.697 & 0.550 & 0.716 & 0.765 & 0.606 \\
Harmony           & emb. & 0.797 & 0.663 & 0.518 & 0.722 & 0.745 & 0.583 \\
fastMNN           & emb. & 0.780 & 0.651 & 0.544 & 0.711 & 0.726 & 0.602 \\
scGPT             & emb. & 0.738 & 0.666 & --    & 0.642 & --    & 0.600 \\
ComBat            & feat. & 0.746 & 0.640 & 0.539 & 0.625 & 0.694 & 0.570 \\
SCALEX            & feat. & 0.702 & 0.574 & 0.510 & 0.598 & --    & 0.531 \\
Scanorama         & feat. & 0.433 & 0.293 & 0.205 & 0.248 & 0.192 & --    \\
Geneformer        & emb. & 0.013 & 0.186 & --    & 0.044 & --    & 0.268 \\
\bottomrule
\end{tabular}
\end{table}

\subsection{ConDo is competitive with deep embedding methods}
\label{sec:results-all}

Table~\ref{tab:all} places ConDo against \emph{all} integration methods, including deep embedding models, on the metric set common to feature and embedding methods. Despite returning corrected expression rather than an optimized embedding, best-of-two ConDo ranks first across all methods on four of six atlases (\textit{DKD}, \textit{Hypomap}, \textit{MousePancreas}, \textit{TabulaSapiens}), second on \textit{ImmuneCell} (behind scANVI), and third on \textit{GTEx} (behind scANVI and scVI). Its most natural comparator is scANVI, the other label-aware method: ConDo exceeds scANVI on four of the six datasets, trailing it only on \textit{GTEx} and (by $0.006$) on \textit{ImmuneCell}---all while producing a corrected gene-expression matrix that an embedding cannot. Unsupervised deep methods (scVI, Harmony) and foundation models (scGPT, Geneformer) are generally outperformed.

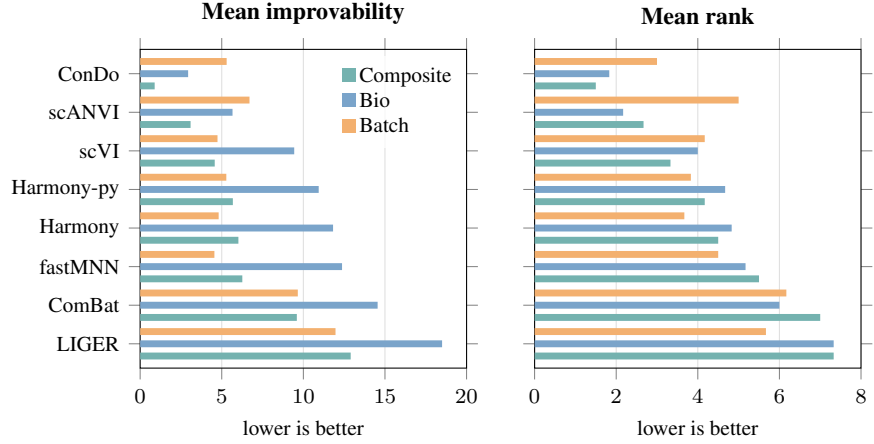
\begin{figure}[t]
\centering
\begin{tikzpicture}
\begin{groupplot}[
  group style={group size=2 by 1, horizontal sep=0.9cm},
  xbar, xmin=0,
  width=5.9cm, height=5.8cm,
  enlarge y limits=0.09,
  symbolic y coords={LIGER,ComBat,fastMNN,Harmony,Harmony-py,scVI,scANVI,ConDo},
  ytick=data,
  tick align=outside,
  xmajorgrids, major grid style={gray!25},
  tick label style={font=\footnotesize},
  label style={font=\footnotesize},
  title style={font=\small\bfseries},
]
\nextgroupplot[title={Mean improvability}, xlabel={lower is better}, xmin=0, xmax=20, bar width=2.6pt]
\addplot[fill=cComposite,draw=none] coordinates {(0.89,ConDo) (3.09,scANVI) (4.57,scVI) (5.68,Harmony-py) (6.02,Harmony) (6.26,fastMNN) (9.60,ComBat) (12.90,LIGER)};
\addplot[fill=cBio,draw=none] coordinates {(2.94,ConDo) (5.66,scANVI) (9.44,scVI) (10.94,Harmony-py) (11.82,Harmony) (12.37,fastMNN) (14.55,ComBat) (18.50,LIGER)};
\addplot[fill=cBatch,draw=none] coordinates {(5.30,ConDo) (6.70,scANVI) (4.74,scVI) (5.28,Harmony-py) (4.81,Harmony) (4.55,fastMNN) (9.66,ComBat) (11.97,LIGER)};
% manual legend (a groupplot \legend accumulates both panels' plots -> 6 swatches)
\node[anchor=north east, fill=white, fill opacity=0.75, text opacity=1, inner sep=3pt, font=\footnotesize]
  at (axis description cs:0.98,0.98) {%
  \begin{tabular}{@{}l@{\,}l@{}}
  \textcolor{cComposite}{\rule{1.4ex}{1.4ex}} & Composite\\
  \textcolor{cBio}{\rule{1.4ex}{1.4ex}} & Bio\\
  \textcolor{cBatch}{\rule{1.4ex}{1.4ex}} & Batch\\
  \end{tabular}};
\nextgroupplot[title={Mean rank}, xlabel={lower is better}, yticklabels={}, xmin=0, xmax=8, bar width=2.6pt]
\addplot[fill=cComposite,draw=none] coordinates {(1.50,ConDo) (2.67,scANVI) (3.33,scVI) (4.17,Harmony-py) (4.50,Harmony) (5.50,fastMNN) (7.00,ComBat) (7.33,LIGER)};
\addplot[fill=cBio,draw=none] coordinates {(1.83,ConDo) (2.17,scANVI) (4.00,scVI) (4.67,Harmony-py) (4.83,Harmony) (5.17,fastMNN) (6.00,ComBat) (7.33,LIGER)};
\addplot[fill=cBatch,draw=none] coordinates {(3.00,ConDo) (5.00,scANVI) (4.17,scVI) (3.83,Harmony-py) (3.67,Harmony) (4.50,fastMNN) (6.17,ComBat) (5.67,LIGER)};
\end{groupplot}
\end{tikzpicture}
\caption{Aggregate performance on the all-methods leaderboard across the six atlases, for the eight methods with full coverage (methods ordered by composite rank, ConDo best-of-two at top). Left: mean improvability (mean per-dataset gap to the best method, $\times100$); right: mean rank. Lower is better for both. Bars are the composite score and its bio-conservation and batch-correction sub-scores. LIGER and Harmony-py are the \texttt{pyliger} and \texttt{harmonypy} implementations.}
\label{fig:agg}
\end{figure}

To better evaluate overall method performance, we further aggregate over the per-dataset scores. 
For each method, we compute its mean \emph{rank} among all methods, and also its mean \emph{improvability}, i.e. the per-dataset gap to the best method ($\times100$). 
The resulting metrics in Figure~\ref{fig:agg} confirms the pattern: over the eight methods with full six-atlas coverage, best-of-two ConDo has the best mean composite rank ($1.50$) and improvability ($0.89$). Its lead is overwhelming on bio-conservation, while it enjoys strong but not leading batch-correction performance.

% Numbers below are reproduced by scripts/aggregate_figure.py --profile all --pgfplots

\subsection{Affine and location-scale capture different batch effects}
\label{sec:results-variants}

The two transforms are complementary rather than redundant (Table~\ref{tab:feature}). Full-affine correction, because it can undo batch effects that mix co-expressed genes, is generally stronger empirically, winning or tying on four of six datasets. Location-scale is decisively better on \textit{ImmuneCell} and \textit{MousePancreas}, where the technical effect is closer to a per-gene shift-and-scale and the extra $O(d^2)$ affine parameters add variance without bias reduction. The best-of-two envelope, which selects the better transform per dataset, is first among feature methods on five of six datasets.

Selecting the transform does not require ground-truth labels: because there are only two options, the choice can be made with the same SCIB-style diagnostics the benchmark already reports (e.g.\ preferring the transform with better batch mixing at equal bio-conservation), or by held-out cells, at the cost of one extra fit. We report the oracle best-of-two to characterize the achievable envelope; in deployment the per-dataset winner can be recovered by this inexpensive two-way model selection.
%A fuller automatic selector is left to future work.

\subsection{Robustness to missing cell-type labels}
\label{sec:results-dropout}

ConDo conditions on cell-type labels, which the benchmark provides but which in deployment would come from imperfect per-batch clustering, leaving some cells unlabeled or low-confidence. We test how ConDo degrades as labels are withheld: on each atlas we hide the cell-type label of a fraction of cells and refit, evaluating two ways of handling the now-unlabeled cells. First, we \emph{exclude} them from the conditional-MMD fit, using only the remaining labeled cells; the unlabeled cells they are still corrected by the fitted map, since $g$ takes features only and so applies to every cell (criterion~(iii)). Second, we pool them into a single ``unknown'' pseudo-type (\emph{bucket}); note that at 100\% of labels hidden, this becomes equivalent to matching marginal distributions with MMD. In both cases, for merge ordering, we use the batch-silhouette criterion, which does not itself depend on cell-type labels.

\begin{figure}[t]
\centering
\includegraphics[trim=0 2mm 0 2mm, clip,width=0.95\linewidth]{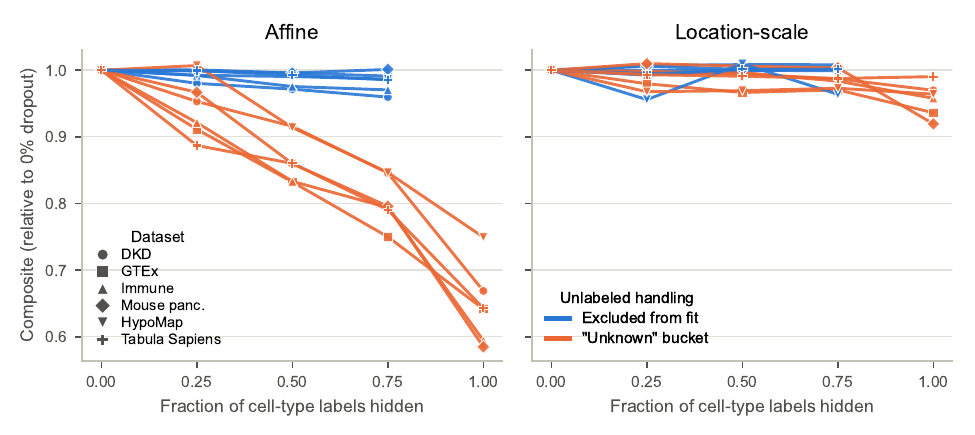}
\caption{Robustness to missing cell-type labels. Each curve is the overall composite relative to the fully-labeled fit as a fraction of cell-type labels is hidden, for the affine (left) and location-scale (right) transforms; color marks how the now-unlabeled cells are handled (excluded from the fit vs.\ pooled into one ``unknown'' bucket) and marker marks the atlas. Excluding unlabeled cells stays within a few percent of baseline for both transforms on all six atlases (and is undefined at $100\%$, where no labeled cells remain to fit); only the affine transform with the ``unknown'' bucket degrades substantially.}
\label{fig:dropout}
\end{figure}

Figure~\ref{fig:dropout} plots the resulting composite score relative to the fully-labeled baseline for each dataset. We first observe that \emph{excluding} unlabeled cells is strikingly robust: across all six atlases and both transforms, hiding up to $75\%$ of labels keeps the composite within about $4\%$ of its fully-labeled value, because the fit simply proceeds on the labeled remainder while $g$ still corrects the rest. Additionally, we see that location-scale is robust under either handling. The practical recipe is therefore to exclude low-confidence cells, under which ConDo tolerates the loss of most of its labels.

\section{Conclusion}
\label{sec:conclusion}
\label{sec:discussion}

We recast single-cell batch integration as confounded domain adaptation and lift ConDo to the many-batch setting via an agglomerative compatibility-graph integrator. On the Open Problems benchmark, our proposed approach is the best feature-space integrator and competitive with deep embedding methods, while uniquely returning corrected expression. The results suggest the embedding-versus-feature tradeoff is less forced than commonly assumed: a label-aware feature method that respects confounding can match or exceed deep embedding methods while returning corrected expressions.

\paragraph{Limitations.} ConDo requires per-cell annotations. In the benchmark these are a provided input, making ConDo comparable to scANVI but not to fully unsupervised methods; in practice they would come from per-batch clustering and annotation, though ConDo tolerates hiding most of these labels. The compatibility graph assumes a single connected component, which need not hold in general. The transformation is linear, so strongly nonlinear batch effects are only partially corrected.

\paragraph{Future Work.} Natural extensions include automatic transform selection, nonlinear conditional matching, and handling compatibility graphs with multiple connected components.

\clearpage
\bibliographystyle{plainnat}
\bibliography{references}

\clearpage
\appendix
\section{Appendix}

\iffalse
\subsection{Related work}
\label{app:related}

Integration methods divide into embedding methods (Harmony \citep{korsunsky2019fast}, scVI/scANVI \citep{lopez2018deep,xu2021probabilistic}, BBKNN \citep{polanski2020bbknn}, foundation models \citep{cui2024scgpt,theodoris2023transfer,rosen2023universal}) and feature methods returning corrected expression (ComBat \citep{johnson2007adjusting}, Scanorama \citep{hie2019efficient}, MNN \citep{haghverdi2018batch}, SCALEX \citep{xiong2022online}). ConDo is a feature method that, unlike the others, matches conditional rather than marginal distributions and is label-aware; scANVI is the only widely used label-aware integrator but produces an embedding. On the methodological side, conditional-distribution matching across domains is studied as generalized label shift and target/conditional shift \citep{tachet2020domain,zhang2013domain}, but those methods reweight samples or optimize a single downstream label rather than producing reusable corrected features. ConDo \citep{mccarter2024towards} learns a feature-space map for general-purpose backward compatibility; we apply it, for the first time at scale, to many-batch single-cell integration via the agglomerative graph construction.
\fi

\subsection{SCIB metric definitions}
\label{app:metrics}

Table~\ref{tab:metricdefs} briefly describes the thirteen SCIB metrics \citep{luecken2022benchmarking} used by the Open Problems Batch Integration Benchmark and combined into the bio-conservation, batch-correction, and overall scores of Section~\ref{sec:setup}. All metrics are computed on the 50-D embedding (a precomputed PCA for embedding methods, or a PCA of the corrected expression for feature methods), min-max scaled across the full method pool, and averaged within each group; the overall score is $0.6\,\text{bio}+0.4\,\text{batch}$.

\begin{table}[h]
\centering
\small
\setlength{\tabcolsep}{4pt}
\caption{The thirteen SCIB metrics grouped by score. \emph{Bio-conservation} rewards preserving biological (cell-type) structure; \emph{batch-correction} rewards mixing cells across batches. Higher is better for all (after scaling). Cell-cycle conservation, isolated-label F1, and kBET are excluded from our scored sets (Section~\ref{sec:setup}).}
\label{tab:metricdefs}
\begin{tabular}{@{}l p{0.72\textwidth}@{}}
\toprule
Metric & Description \\
\midrule
\multicolumn{2}{@{}l}{\emph{Bio-conservation}}\\
Cell-type ASW & Average silhouette width of cells grouped by cell-type label: how well cell types separate in the embedding. \\
NMI & Normalized mutual information between graph-based (Leiden) clusters and cell-type labels. \\
ARI & Adjusted Rand index between the same clusters and cell-type labels. \\
Isolated-label ASW & Silhouette width restricted to cell types present in few batches, testing whether rare/isolated labels are preserved. \\
Isolated-label F1 & F1 of recovering isolated-label cells by clustering, a second view of rare-label preservation. \\
cLISI & Cell-type local inverse Simpson's index: neighborhoods should be homogeneous in cell type (not mixed). \\
HVG overlap & Overlap of highly variable genes before vs.\ after integration, penalizing loss of biological signal (feature output only). \\
Cell-cycle conservation & Retention of cell-cycle-score variance after integration. \\
\midrule
\multicolumn{2}{@{}l}{\emph{Batch-correction}}\\
Batch ASW & Silhouette width of cells grouped by batch, rescaled so that well-mixed (low-silhouette) batches score high. \\
Graph connectivity & Whether cells of the same type form a connected kNN subgraph across batches. \\
PCR & Principal-component regression: reduction in the variance explained by batch after integration. \\
iLISI & Integration local inverse Simpson's index: neighborhoods should be well mixed across batches. \\
kBET & $k$-nearest-neighbor batch-effect test: local batch composition should match the global composition. \\
\bottomrule
\end{tabular}
\end{table}

\subsection{Bio-conservation and batch-correction decomposition}
\label{app:decomp}

Tables~\ref{tab:bio} and~\ref{tab:batch} decompose the feature-leaderboard overall score (Table~\ref{tab:feature}) into its bio-conservation and batch-correction components (the SCIB sub-scores that the $0.6\,\text{bio}+0.4\,\text{batch}$ composite averages). As in the main tables, each ConDo variant is scored within its own method pool. Two patterns stand out. First, ConDo's advantage over ComBat is concentrated in \emph{bio-conservation}: ConDo attains the highest bio score on five of six datasets, often by a wide margin (e.g.\ \textit{Hypomap} $0.753$ vs.\ $0.524$, \textit{MousePancreas} $0.858$ vs.\ $0.692$), whereas the two methods are much closer on batch correction, where ComBat is sometimes ahead. This is the expected signature of conditioning on cell type: it preserves biological structure rather than buying batch mixing at its expense. Second, the affine and location-scale variants differ mainly in \emph{how they trade bio against batch}: location-scale raises batch mixing on some datasets (e.g.\ \textit{DKD} $0.786$ vs.\ affine $0.714$) but can sacrifice bio-conservation badly on others (\textit{TabulaSapiens} bio $0.623$ vs.\ affine $0.805$), explaining why neither transform dominates and why selecting between them per dataset helps.

\begin{table}[h]
\centering
\small
\setlength{\tabcolsep}{4pt}
\caption{Bio-conservation sub-score per dataset (feature leaderboard; higher is better). \textbf{Bold} marks the best method per dataset.}
\label{tab:bio}
\begin{tabular}{lcccccc}
\toprule
Method & DKD & GTEx & Hypomap & Immune & Mouse & TabSap \\
\midrule
ConDo (best-of-two)    & \textbf{0.883} & 0.698 & 0.753 & \textbf{0.860} & \textbf{0.858} & \textbf{0.805} \\
\quad ConDo affine     & 0.883 & 0.698 & 0.753 & 0.879 & 0.736 & 0.805 \\
\quad ConDo loc.-scale & 0.818 & 0.664 & \textbf{0.777} & 0.860 & 0.858 & 0.623 \\
ComBat                     & 0.773 & \textbf{0.798} & 0.524 & 0.805 & 0.692 & 0.774 \\
\bottomrule
\end{tabular}
\end{table}

\begin{table}[h]
\centering
\small
\setlength{\tabcolsep}{4pt}
\caption{Batch-correction sub-score per dataset (feature leaderboard; higher is better). \textbf{Bold} marks the best method per dataset.}
\label{tab:batch}
\begin{tabular}{lcccccc}
\toprule
Method & DKD & GTEx & Hypomap & Immune & Mouse & TabSap \\
\midrule
ConDo (best-of-two)    & 0.714 & \textbf{0.568} & 0.658 & \textbf{0.727} & 0.666 & \textbf{0.548} \\
\quad ConDo affine     & 0.714 & 0.568 & 0.658 & 0.668 & 0.571 & 0.548 \\
\quad ConDo loc.-scale & \textbf{0.786} & 0.391 & 0.532 & 0.727 & 0.666 & 0.356 \\
ComBat                     & 0.749 & 0.565 & \textbf{0.747} & 0.629 & \textbf{0.763} & 0.520 \\
\bottomrule
\end{tabular}
\end{table}

Tables~\ref{tab:bioall} and~\ref{tab:batchall} give the same decomposition for the all-methods leaderboard (Table~\ref{tab:all}), where ConDo is scored against the embedding methods on the 7 shared metrics. The picture is consistent with the feature comparison: ConDo's edge is in \emph{bio-conservation}---best-of-two ConDo has the top bio score on \textit{DKD} and \textit{Hypomap} and is within $0.003$ of scANVI on \textit{MousePancreas}---while on \emph{batch-correction} the embedding methods, which optimize an embedding directly for mixing, are often ahead. The affine/location-scale split again reflects a bio-versus-batch tradeoff (e.g.\ on \textit{ImmuneCell} location-scale buys the best batch score, $0.790$, at some cost to bio).

\begin{table}[h]
\centering
\small
\setlength{\tabcolsep}{3.5pt}
\caption{Bio-conservation sub-score per dataset (all-methods leaderboard; higher is better). \textbf{Bold} marks the best method per dataset; $^\ast$ marks label-aware methods.}
\label{tab:bioall}
\begin{tabular}{llcccccc}
\toprule
Method & type & DKD & GTEx & Hypomap & Immune & Mouse & TabSap \\
\midrule
ConDo (best-of-two)         & feat$^\ast$ & \textbf{0.842} & 0.686 & \textbf{0.751} & 0.750 & 0.820 & 0.655 \\
\quad ConDo affine          &             & 0.842 & 0.686 & 0.751 & 0.750 & 0.761 & 0.655 \\
\quad ConDo location-scale  &             & 0.753 & 0.585 & 0.557 & 0.723 & 0.820 & 0.519 \\
scANVI            & emb$^\ast$ & 0.769 & \textbf{0.797} & 0.484 & \textbf{0.796} & \textbf{0.823} & \textbf{0.671} \\
scVI              & emb. & 0.737 & 0.723 & 0.512 & 0.713 & 0.807 & 0.622 \\
Harmony           & emb. & 0.758 & 0.641 & 0.531 & 0.689 & 0.791 & 0.562 \\
fastMNN           & emb. & 0.765 & 0.654 & 0.519 & 0.674 & 0.746 & 0.580 \\
scGPT             & emb. & 0.721 & 0.657 & --    & 0.607 & --    & 0.596 \\
ComBat            & feat. & 0.750 & 0.673 & 0.480 & 0.621 & 0.697 & 0.586 \\
SCALEX            & feat. & 0.644 & 0.585 & 0.531 & 0.579 & --    & 0.537 \\
Scanorama         & feat. & 0.281 & 0.185 & 0.140 & 0.094 & 0.037 & --    \\
Geneformer        & emb. & 0.022 & 0.303 & --    & 0.073 & --    & 0.428 \\
\bottomrule
\end{tabular}
\end{table}

\begin{table}[h]
\centering
\small
\setlength{\tabcolsep}{3.5pt}
\caption{Batch-correction sub-score per dataset (all-methods leaderboard; higher is better). \textbf{Bold} marks the best method per dataset; $^\ast$ marks label-aware methods.}
\label{tab:batchall}
\begin{tabular}{llcccccc}
\toprule
Method & type & DKD & GTEx & Hypomap & Immune & Mouse & TabSap \\
\midrule
ConDo (best-of-two)         & feat$^\ast$ & 0.813 & \textbf{0.707} & 0.403 & 0.777 & 0.695 & \textbf{0.685} \\
\quad ConDo affine          &             & 0.813 & 0.707 & 0.403 & 0.777 & 0.610 & 0.685 \\
\quad ConDo location-scale  &             & 0.837 & 0.629 & 0.510 & \textbf{0.790} & 0.695 & 0.542 \\
scANVI            & emb$^\ast$ & 0.849 & 0.660 & 0.520 & 0.721 & 0.672 & 0.576 \\
scVI              & emb. & 0.846 & 0.659 & 0.608 & 0.720 & \textbf{0.701} & 0.581 \\
Harmony           & emb. & \textbf{0.855} & 0.695 & 0.498 & 0.772 & 0.676 & 0.614 \\
fastMNN           & emb. & 0.803 & 0.645 & 0.581 & 0.766 & 0.695 & 0.636 \\
scGPT             & emb. & 0.763 & 0.679 & --    & 0.695 & --    & 0.606 \\
ComBat            & feat. & 0.740 & 0.589 & \textbf{0.627} & 0.630 & 0.690 & 0.544 \\
SCALEX            & feat. & 0.790 & 0.558 & 0.479 & 0.626 & --    & 0.522 \\
Scanorama         & feat. & 0.661 & 0.454 & 0.301 & 0.480 & 0.424 & --    \\
Geneformer        & emb. & 0.000 & 0.009 & --    & 0.000 & --    & 0.028 \\
\bottomrule
\end{tabular}
\end{table}

\subsection{Sensitivity to the agglomerative merge order}
\label{app:ordering}

\begin{figure}[tbp]
\centering
\includegraphics[width=0.92\linewidth]{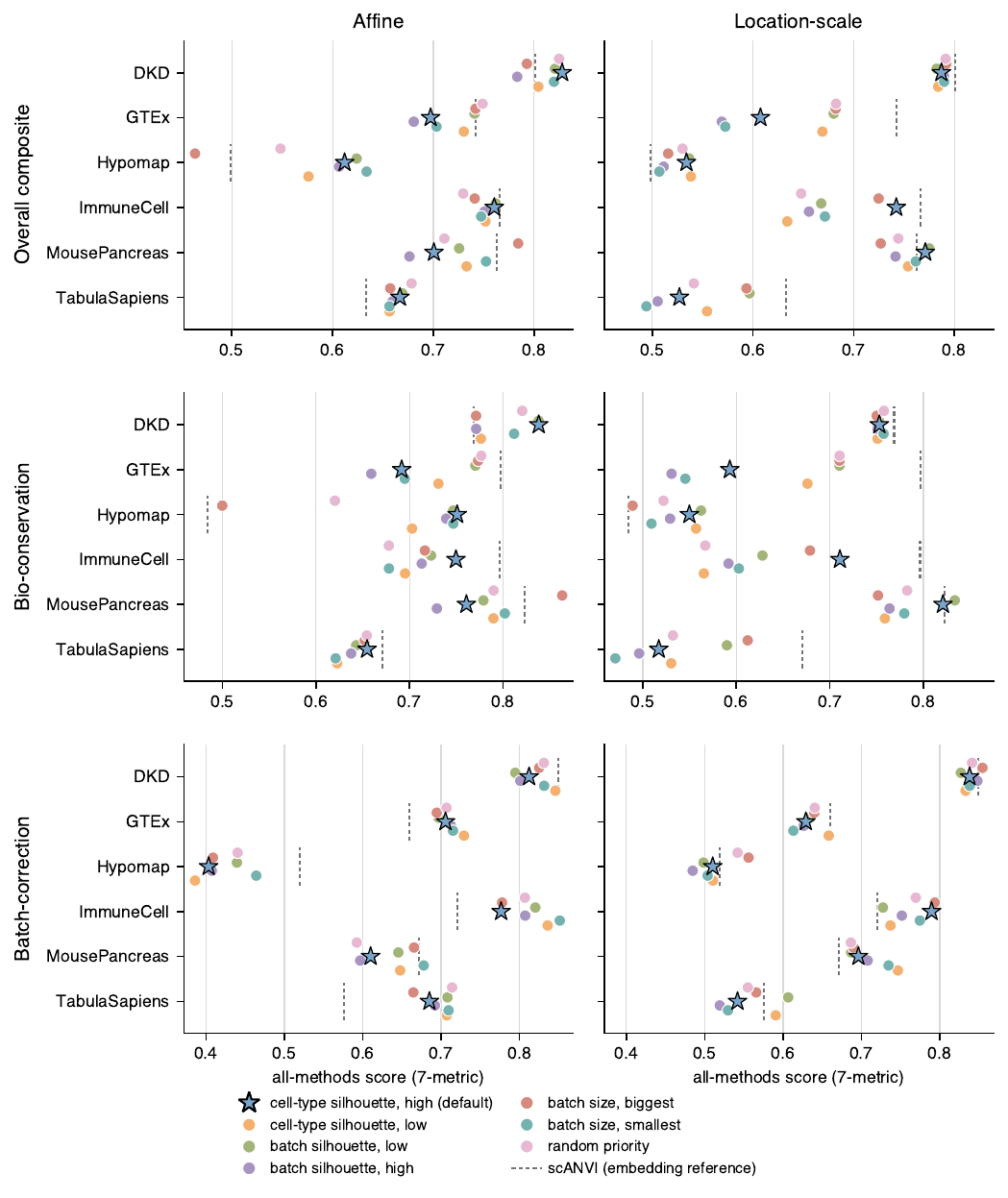}
\caption{Merge-order ablation across seven seed/merge-order ranking strategies, under the affine (left) and location-scale (right) transforms, decomposed into the overall composite (top row), bio-conservation (middle), and batch-correction (bottom) scores---all on the all-methods 7-metric set, as in Table~\ref{tab:all}. Each dot is one strategy, colored by strategy; the emphasized star is the deployed cell-type-silhouette default; the dashed tick marks scANVI, the strongest embedding comparator, on that dataset. Horizontal spread within a dataset shows how much the merge order matters---small on most atlases, wider on the many-batch \textit{Hypomap} and \textit{MousePancreas}, and concentrated in bio-conservation. The default is among the strongest strategies on bio-conservation but the weakest on batch mixing, so its composite competitiveness comes from the $0.6$-weighted bio term; nearly all strategies sit at or right of the scANVI tick on the composite, so ConDo's standing is robust to the order. Due to an oversight, the location-scale panels use weight decay $10^{-5}$, so the composite score for the default merge order differs by ${\le}0.008$ from the $10^{-4}$ location-scale result reported in the main table (Table~\ref{tab:all}).}
\label{fig:ordering}
\end{figure}

The agglomerative integrator of Section~\ref{sec:method-agg} scores each batch by its within-batch cell-type silhouette and uses that score for two purposes: choosing the seed and ordering the greedy merges. Because the final result depends on this order, we ask how much the choice of scoring rule matters. We rerun the full pipeline with seven ranking strategies, changing \emph{only} the score that drives the seed and merge order and holding every other hyperparameter fixed at the main-text configuration. Besides the deployed cell-type silhouette (highest first), we test its mirror (lowest first); the per-batch \emph{batch} silhouette in both directions (lowest first merges already-well-mixed batches earliest); batch size in both directions; and a fixed random priority. We run each strategy for both the affine and location-scale transforms, and score them on the all-methods 7-metric leaderboard exactly as in Table~\ref{tab:all}, reporting the overall composite together with its bio-conservation and batch-correction components (Figure~\ref{fig:ordering}). Strategy-versus-strategy comparisons within a dataset are computed on the same machine and are internally consistent; the small cross-machine drift noted in the main text affects only cross-dataset absolute values.

Three observations follow. First, the merge order has a real but bounded effect: across the seven strategies the overall score spans at most $0.03$--$0.07$ on four of the six atlases, widening to $0.11$ (\textit{MousePancreas}) and $0.17$ (\textit{Hypomap}), the two atlases with the most batches ($56$ and $24$), where ordering choices compound. Second, no strategy dominates, and the deployed cell-type-silhouette rule is competitive but not uniformly best: on the single-transform means it sits mid-pack, with the batch-silhouette-low rule slightly ahead ($0.724$ vs.\ $0.711$ affine; $0.673$ vs.\ $0.661$ location-scale). Third, and most important for our claims, this single-transform gap all but vanishes under the best-of-two envelope that Table~\ref{tab:all} actually reports: batch-silhouette-low averages $0.732$ against the deployed rule's $0.723$ on the all-methods composite, and the two are tied at $0.744$ on the feature leaderboard. We therefore retain the cell-type-silhouette rule as the headline configuration---it carries the same label-aware motivation as the rest of the method---and read this ablation as evidence that ConDo's standing is \emph{robust} to the merge order rather than as a case for tuning it: every strategy, including random priority, leaves ConDo well ahead of the other feature methods and competitive with the deep embedding models. Selecting a strategy post hoc on these same six atlases would moreover be exactly the benchmark-tuning we caution against in Section~\ref{sec:intro}.

Decomposing the composite into its bio-conservation and batch-correction components (Figure~\ref{fig:ordering}, middle and bottom rows) shows where this behavior originates. The cell-type-silhouette default is among the strongest strategies on \emph{bio-conservation}---best of the seven on four of six atlases and second on average---which is the expected consequence of seeding and ordering by how cleanly cell types separate within each batch. It pays for this with the weakest \emph{batch-correction} of any strategy on average, since it never optimizes for batch mixing; strategies that merge small or already-well-mixed batches first (\emph{batch size, smallest}; \emph{batch silhouette, low}) recover more batch mixing instead. Because the composite weights bio-conservation at $0.6$, the default's biology-first behavior keeps it composite-competitive despite this batch deficit. The order-sensitivity itself is likewise concentrated in bio-conservation (spread up to $0.25$ on \textit{Hypomap}) rather than batch-correction (spread $\le 0.09$ on every atlas), so the wide composite swings on the many-batch atlases are a bio-conservation phenomenon. This is consistent with the paper's central claim: what the merge order chiefly governs is how much biological structure the conditional matching preserves.

\subsection{ConDo fit runtime}
\label{app:runtime}

Table~\ref{tab:condo-fit-runtime} reports the wall-clock fit time of ConDo on each atlas for the two transforms under the shared official configuration. Runtime grows with dataset size, and the dense affine map is not uniformly slower than location-scale: the two are within a factor of two on every atlas, so the affine variant's $O(d^2)$ parameters (Section~\ref{sec:bg-transform}) do not impose a prohibitive time cost at these scales.

\begin{table}[t]
\centering
\caption{ConDo fit runtime (minutes) on each benchmark dataset, for the
affine and location-scale transforms under the shared official configuration
(MMD, features, $n_{\text{epochs}}{=}5$, weight decay $10^{-4}$,
\texttt{celltype\_silhouette} ordering). Fit time is wall-clock for the
complete per-dataset fit (read, agglomerative integration, write) on one
NVIDIA RTX PRO 6000 GPU. Wall-clock includes the gzip write of the corrected
output, which accounts for at most ${\sim}11\%$ of fit time (measured
$4.6\%$/$4.8\%$/$11.4\%$ on DKD/GTEx/HypoMap); input reading is negligible
($<3$\,s), so these times are within ${\sim}10\%$ of pure integration compute.}
\label{tab:condo-fit-runtime}
\begin{tabular}{lrrr}
\toprule
Dataset & Cells & Affine (min) & Location-scale (min) \\
\midrule
DKD & 39k & 6.8 & 4.8 \\
GTEx v9 & 209k & 33.4 & 25.4 \\
Immune cell atlas & 330k & 50.5 & 60.1 \\
Mouse pancreas & 302k & 58.1 & 52.2 \\
HypoMap & 385k & 26.3 & 51.6 \\
Tabula Sapiens & 483k & 94.5 & 110.2 \\
\midrule
Total & & 269.6 & 304.3 \\
\bottomrule
\end{tabular}
\end{table}

\end{document}